\documentclass[onecolumn,preprintnumbers,amsmath,amssymb,superscriptaddress]{revtex4}
\usepackage{graphicx,subfigure,epsfig,epstopdf}
\usepackage{graphicx}% Include figure files
\usepackage{dcolumn}% Align table columns on decimal point
\usepackage{bm}% bold math
\usepackage{amsmath,amssymb,upgreek}
\input{epsf}
\input{epsfx}
\makeatletter

\newcommand{\Rmnum}[1]{\expandafter\@slowromancap\romannumeral #1@}
\makeatother

\begin{document}
\title{Controlling the Spontaneous Emission Rate of Monolayer MoS$_2$ in a Photonic Crystal Nanocavity}

\date{Aug 25, 2013}

\author{Xuetao Gan}
\affiliation{School of Science, Northwestern Polytechnical University, Xi'an 710072, China}
\affiliation{Department of Electrical Engineering, Columbia University, New York, NY 10027, USA}
\author{Yuanda Gao}
\affiliation{Department of Mechanical Engineering, Columbia University, New York, NY 10027, USA}
\author{Kin Fai Mak}
\affiliation{Department of Physics, Columbia University, New York, NY 10027, USA}
\author{Xinwen Yao}
\affiliation{Department of Electrical Engineering, Columbia University, New York, NY 10027, USA}
\author{Ren-Jye Shiue}
\affiliation{Department of Electrical Engineering and Computer Science, Massachusetts Institute of Technology, Cambridge, MA 02139, USA}
\author{Arend van der Zande}
\affiliation{Department of Mechanical Engineering, Columbia University, New York, NY 10027, USA}
\author{Matthew Trusheim}
\affiliation{Department of Electrical Engineering and Computer Science, Massachusetts Institute of Technology, Cambridge, MA 02139, USA}
\author{Fariba Hatami}
\affiliation{Department of Physics, Humboldt-Universit\"at zu Berlin, Newtonstrasse 15, 12489 Berlin, Germany}
\author{Tony F. Heinz}
\affiliation{Department of Electrical Engineering, Columbia University, New York, NY 10027, USA}
\affiliation{Department of Physics, Columbia University, New York, NY 10027, USA}
\author{James Hone}
\affiliation{Department of Mechanical Engineering, Columbia University, New York, NY 10027, USA}
\author{Dirk Englund}
\email{englund@mit.edu}
\affiliation{Department of Electrical Engineering and Computer Science, Massachusetts Institute of Technology, Cambridge, MA 02139, USA}

%\section{Introduction}
\begin{abstract}
We report on controlling the spontaneous emission (SE) rate of a molybdenum disulfide (MoS$_2$)  monolayer coupled with a planar photonic crystal (PPC) nanocavity. Spatially resolved photoluminescence (PL) mapping shows strong variations of emission when the MoS$_2$ monolayer is on the PPC cavity, on the PPC lattice, on the air gap, and on the unpatterned gallium phosphide substrate. Polarization dependences of the cavity-coupled MoS$_2$ emission show a more than 5 times stronger extracted PL intensity than the un-coupled emission, which indicates an underlying cavity mode Purcell enhancement of MoS$_2$ SE rate exceeding a factor of 70. 
\end{abstract}
%\pacs{42.50.Ct, 42.50.Dv, 42.70.Qs, 78.67.Hc}
\maketitle

%\section{Introduction}

The recent finding that a single atomic layer of transition metal dichalcogenides (TMDs) can exhibit a large, direct bandgap~\cite{Wang2012o,Mak2010b,Han2011a,Kadantsev2012} opens the possibility of a new range of atomically thin materials for electronic and electro-optic devices.  Monolayer molybdenum disulfide (MoS$_2$) has been used to fabricate field-effect transistors (FETs) with a  carrier-mobility of 200 cm$^2$V$^{-1}$s$^{-1}$ and On/Off ratios exceeding 10$^8$ at room temperature, comparable to  those obtained in graphene nanoribbon-based FETs~\cite{Radisavljevic2011}. Optical studies have shown that monolayer MoS$_2$ exhibits a photoluminescence (PL) quantum yield that is enhanced by a factor more than 10$^4$ compared with the bulk crystal~\cite{Mak2010b,Splendiani2010}. However, the PL efficiency of monolayer MoS$_2$ is still very low at $\sim10^{-2}$ because the nonradiative recombination rate $1/\tau_{nr}$ far exceeds the spontaneous emission (SE) rate $1/\tau_r$~\cite{Mak2010b}. For MoS$_2$ monolayers on SiO$_2$ substrate, values of $\tau_{nr}\sim 100$ ps and $\tau_{r}\sim 10$ ns were estimated at room temperature~\cite{Korn2012,Mak2010b}. Here, we show that the SE efficiency of an MoS$_2$ monolayer can be greatly enhanced by exploiting the strong Purcell effect in photonic crystal nanocavities to shorten the radiative recombination time. After depositing an MoS$_2$ monolayer onto  a planar photonic crystal (PPC) nanocavity, we observe an enhancement of the external extracted PL intensity by a factor of  5.4 above the background. This strong enhancement exists even though the collection is from the sub-wavelength cavity mode and the surrounding focal spot region. Taking into account this spatial averaging, we deduce that the SE rate enhancement into the cavity mode corresponds to nearly a factor of 70, in close agreement with theory.  These results indicate that by exploiting the strong Purcell effect in optical cavities with wavelength-scale mode volume and high quality ($Q$) factor, it is possible to achieve roughly two orders of magnitude improvement in the MoS$_2$ PL efficiency. This gain opens the door to efficient light emissions from, and strong light-matter interactions with, materials of atomic thickness. 

 \begin{figure}[th!]\centering
\includegraphics[width=6in]{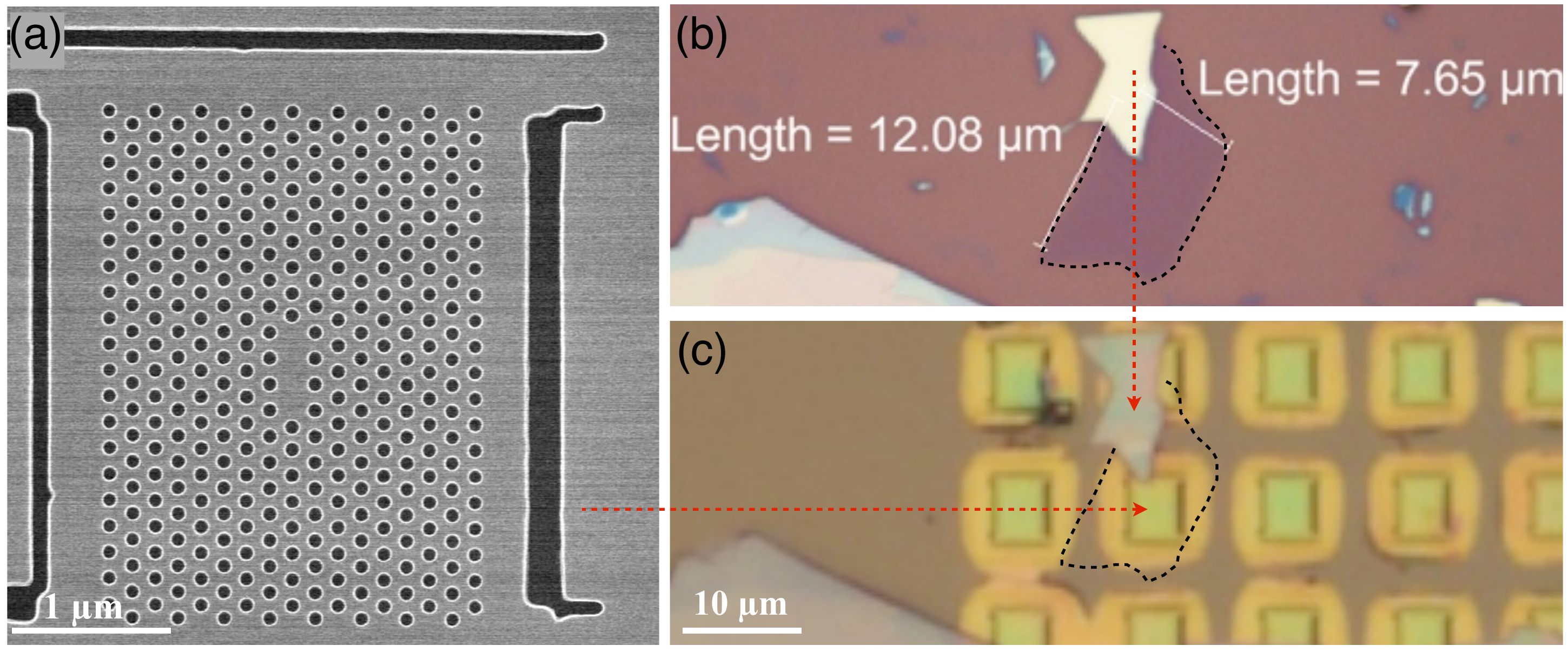}
\caption{{\small (Color online) (a) SEM image of the L3 PPC nanocavity before the transfer of  MoS$_2$ monolayer. (b) Optical microscope image of the exfoliated MoS$_2$ film on a polymeric sacrificial substrate. The monolayer is shown in the purple region indicated by the dashed black line. (c) Optical microscope image of a finished device. The single-layer MoS$_2$ is not visible, but its overlap with the  PPC cavity is verified by the above multi-layer MoS$_2$ flake and by the fluorescence mapping image shown in Fig. 2(a). }}
\label{fig:equipment_layout} 
\end{figure}
\vspace{12pt}

The experiment employs PPC nanocavities fabricated in a 138~nm thick gallium phosphide (GaP) membrane using electron-beam lithography, dry etching, and wet chemical undercutting of an AlGaP sacrificial layer~\cite{Gan2012b}. The cavity design is a linear three-missing hole (L3) defect~\cite{Noda2003Nature} with a lattice spacing $a=165$~nm and an air-hole radius $r=0.3a$, yielding resonant modes in the wavelength range of 600~nm-700~nm to overlap the PL spectrum of the monolayer MoS$_2$. Figure 1(a) shows a scanning electron microscope (SEM) image of the PPC nanocavity before the deposition of MoS$_2$. Trenches around the PPC lattice aid in the removal of the sacrificial layer in a  hydrofluoric acid bath. The monolayer MoS$_2$ is prepared by mechanical exfoliation onto a polymeric sacrificial substrate, as shown in the optical microscope image in Fig. 1(b). Due to the optical interference, MoS$_2$ monolayer is clearly visible in the purple region indicated by the dashed black line, which is also confirmed by a micro-Raman spectroscopy~\cite{Mos2010}. The MoS$_2$ sheet is then transferred onto PPC nanocavities through a precision transfer technique with the help of the polymeric sacrificial substrate, which is removed from the final device by high-temperature annealing~\cite{2010.NNano.Hone.graphene}. Figure 1(c) shows the finished device. An PPC nanocavity is covered uniformly by the MoS$_2$ monolayer, which is clearly distinguished by correlating the above multi-layer MoS$_2$ flake.  

 \begin{figure}[th!]\centering
\includegraphics[width=6in]{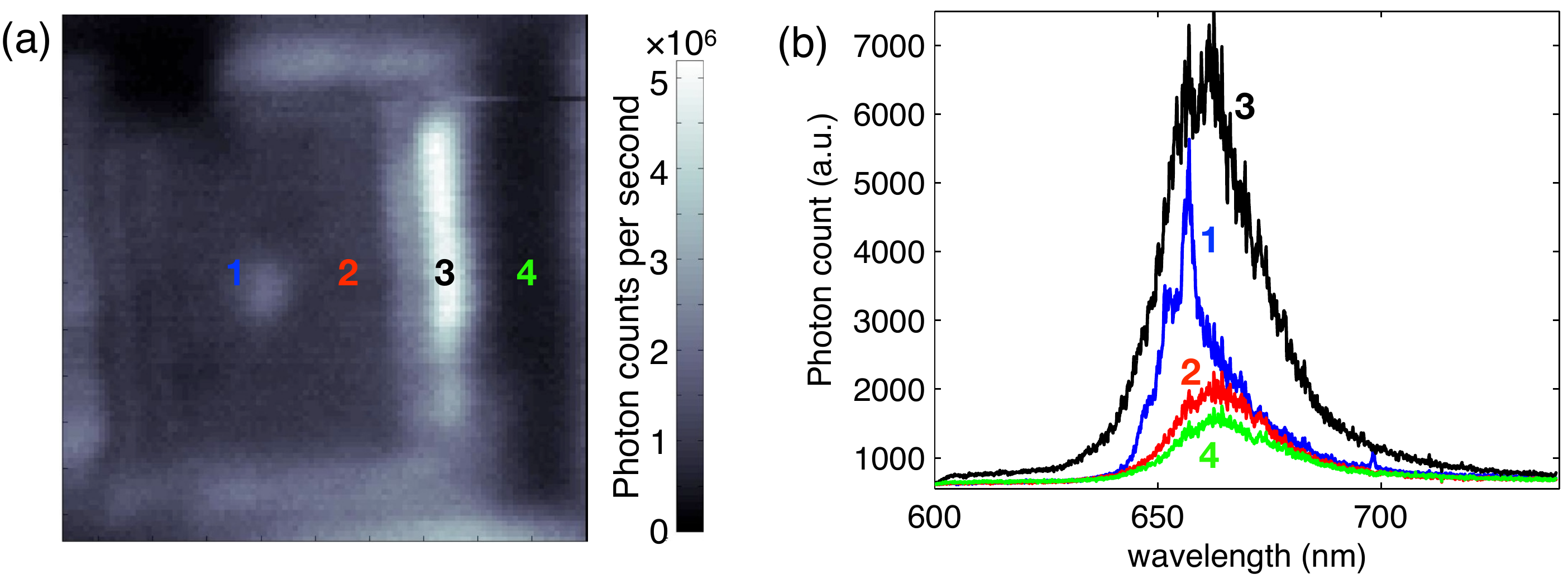}
\caption{{\small (Color online)  (a)  Micro-PL spatial mapping of the device, showing four individual emission profiles. (b) PL spectra collected from the four different locations on the sample.}}
\label{fig:equipment_layout}
\end{figure}
\vspace{12pt}

We characterize the device on a micro-PL confocal microscope with a 532~nm continuous-wave excitation laser, focused to a beam diameter of $\sim$400 nm and with a power of $\sim$50 $\upmu$W. To study the modifications on the  MoS$_2$ SE, we spatially scan the device in 50 nm steps on a piezo stage and detect the MoS$_2$ PL using an avalanche photodiode. Figure 2(a) shows the spatially resolved PL. By correlating it with the SEM image shown in Fig. 1(a), we observe four individual emission profiles of MoS$_2$ due to different substrates, as marked in Fig. 2(a). The spectrally resolved  PL of the four regions are shown in Fig. 2(b).  The result reveals that the PL collected from region 3, where the MoS$_2$ sheet is suspended over a 300~nm  wide trench, is significantly brighter than that obtained from region 4 on the bulk GaP membrane. This is expected due to the suppression of the PL quantum yield by the substrate~\cite{Mak2010b} and the total internal reflection of the high-index GaP slab~\cite{2005.Science.Noda.SE_control}, which sharply reduces the PL collection efficiency. On both regions, the monolayer MoS$_2$ emits the same fluorescence spectrum centered around 660~nm due to the direct electronic bandgap~\cite{Mak2010b}. 

On the PPC, we observe  both an enhancement and a suppression of the MoS$_2$ PL emission.  In region 2, due to the coupling between the periodic air-holes of the PPC lattice and the MoS$_2$ sheet, the in-plane emission channel is inhibited by the in-plane photonic bandgap, which overlaps with the emission band of the monolayer MoS$_2$. Therefore, the SE should be re-directed into near-vertical $k$-vectors within the PPC light cone~\cite{2005.Science.Noda.SE_control}. This SE redistribution and the higher collection efficiency  from the PPC lattice enhances the collection of emission into the vertical direction via the suppression of emission into in-plane PPC modes.   Hence, the collected photon flux from region 2 is brighter than that from the bulk GaP membrane, as confirmed from the PL spectra. However, the PL collected from the L3 defect (region 1) shows even brighter emission than that from region 2. Comparing the spectra acquired from region 1 and region 2, it is clear that this enhancement mainly results from a greatly amplified photon flux of the two peaks at the wavelengths of 655.4~nm and 656.9~nm. 
The polarization dependences of the two peaks from region 1 are then resolved by rotating a polarizer in the PL collection path of the microscope setup. The obtained spectra are shown in Fig. 3(a), where $\phi$ denotes the angle between the cavity $y$-axis and the polarization direction of the polarizer. These spectra indicate the two peaks at 655.4~nm and 656.9~nm are the resonant modes of the L3 cavity with expected polarization and wavelength dependences matching the three-dimensional finite difference time domain (FDTD) simulations~\cite{2007.APL.UK_ppl.mode_structure_L3,Schwagmann2012}, which also confirm other resonant modes of the peaks at longer wavelength. Therefore, over the cavity defect region, the simultaneous suppression of SE into in-plane PPC modes together with the cavity mode Purcell enhancement of SE rate results in a dramatic reshaping of the MoS$_2$ SE, as was previously shown for single emitters~\cite{2005.PRL.Englund,Kress2005,Kaniber2007} and quantum wells~\cite{2005.Science.Noda.SE_control} in PPC cavities.  

For simplicity, we designate the two resonant modes at 655.4~nm and 656.9~nm as mode 1 and mode 2, respectively.  Fitting the  peaks to Lorentzian lineshapes, we find that the $Q$ factors of the two modes are $Q_0=220$ and 320, respectively, which degrade from the initial $Q$ factors of 880 and 800 of the unloaded cavity due to the spectrally overlap with the absorption resonance of the monolayer MoS$_2$~\cite{Mak2010b,Gan2012,Gan2013b}. The simulated cavity fields of modes 1 and 2 are shown in the inset of Fig. 3(a), which have mode volumes ($V_{mode}$) of 0.63 and 0.33$(\lambda/n)^3$, where $n$ is the refractive index of GaP. 

 \begin{figure}[th!]\centering
\includegraphics[width=6.5in]{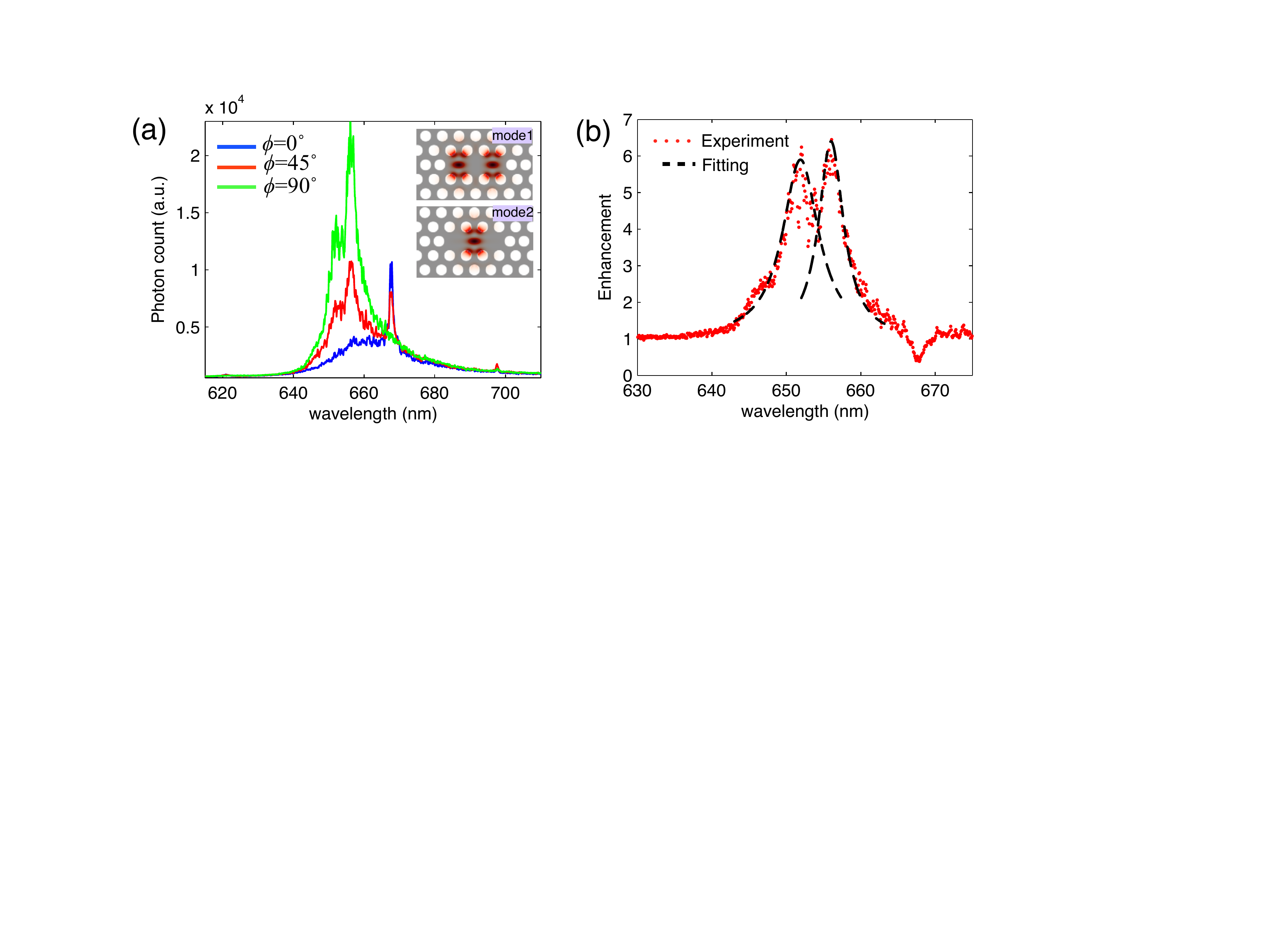}
\caption{{\small (Color online) (a) Polarization dependences of  the cavity-coupled MoS$_2$ PL spectra, where $\phi$ denotes the angle between the cavity $y$-axis and the polarization direction of the polarizer. Inset: Simulated field distributions of resonant modes at 655.4~nm and 656.9~nm. (b) Spectrally resolved enhancement of PL emission calculated from the PL spectra shown in (a) with $\phi=90^\circ$ and $0^\circ$ (dotted line), which can be well fitted to a theoretical model considering the SE rate modifications and coupling efficiencies (dashed line). }}
\label{fig:equipment_layout}
\end{figure}
\vspace{12pt}
% DIRK to Xuetao: Note that in the next paragraph, we first have to make the argument for why a higher Purcell effect results in greater brightness. In the unsaturated driving limit, this is because we enhance a poor emitter with an internal quantum efficiency of $\beta\ll 1$ to $\beta F$

To quantitatively anaylse the cavity enhancement of MoS$_2$ SE rate, we model the coupled MoS$_2$-cavity system by considering the MoS$_2$ as a collection of excitonic dipole emitters.  The exciton recombination rate is given by a sum over radiative  and non-radiative recombination rates, $\Gamma=\Gamma_{r}+\Gamma_{nr}$. In our experiments, the PL intensities at the resonant wavelengths show linear dependence on the excitation power, verifying that the SE processes are far below the saturation rate of the MoS$_2$ sheet. Therefore, the emission power $P$ is proportional to $P_{in}A \Gamma_{r}/(\Gamma_{r}+\Gamma_{nr}), $ where $P_{in}$ is the excitation power and $A$ is the absorbance of monolayer MoS$_2$ at the excitation wavelength. Because $\Gamma_{nr}\gg \Gamma_{r}$ in MoS$_2$ and the finite collection angle of optics, we can approximate for all of our experiments that $P\propto  \eta \Gamma_{r}/\Gamma_{nr} $, where $\eta$ is the collection efficiency of the PL emission. Here, we consider the excitons as an ensemble of emitters $\mathbf{\upmu}$ in the MoS$_2$ on a bulk substrate have a natural SE rate $\Gamma_0(\lambda)\mathrm{d}\lambda$ with a transition rate corresponding to the spectral range from $\lambda$ to $\lambda+\mathrm{d}\lambda$. The modified SE distribution when the MoS$_2$ sheet is on the PPC nanocavity is given by 
\begin{equation}
\Gamma(\lambda)\mathrm{d}\lambda = \Gamma_0(\lambda)\mathrm{d}\lambda[F_{c,0}L(\lambda) |\psi|^{2} +F_{PC}]
\end{equation}
\noindent 
Here, $L(\lambda)=1/[1+4Q^2(\frac{\lambda}{\lambda_{c,0}}-1)^2]$ denotes the cavity's Lorentzian spectrum with $\lambda_{c,0}$ as the resonant wavelength, and $\psi=\mathbf{E}\cdot\mathbf{\upmu}/|\mathbf{E}_{max}||\mathbf{\upmu}|$ denotes the  spatial and angular overlaps between the emitter dipole $\mathbf{\upmu}$ and the cavity field $\mathbf{E}$. The factor $F_{c,0}=\frac{3}{4\pi^{2}}\frac{Q}{V_{mode}}(\frac{\lambda_{c,0}}{n})^3$ is the maximum SE enhancement (Purcell) factor of the cavity mode when the emitter dipole  is on resonance with the cavity and spatially aligned with the cavity field. The term $F_{PC}$ accounts for the suppression of the SE rate by the PPC lattice and modes other than the cavity mode \cite{2005.PRL.Englund,2005.Science.Noda.SE_control}. 

 The total cavity-coupled MoS$_2$ emission spectrum $I_\mathrm{\phi}(\lambda)$ with different polarizations $\phi$ can be  fitted to a model that considers both the SE rate modifications and the collection efficiencies of the cavity mode and averaged PPC leaky modes.  We calculate $I_\mathrm{\phi}(\lambda)$ by integrating the SE rate given in Eq. (1) over the spatial and in-plane orientation densities of the emitter dipoles $\rho(\mathbf{r}, \lambda, \mathbf{\upmu})$
\begin{equation}
I_\mathrm{\phi}(\lambda)=\Gamma_0(\lambda)\int\mathrm{d}\mathbf{\upmu}\mathrm{d}^2\mathbf{r}[\eta_{c,0}F_{c,0}L(\lambda) |\psi|^{2}\sin(\phi)+\eta_{PC}F_{PC}]\rho(\mathbf{r}, \lambda, \mathbf{\upmu})
\end{equation}
\noindent Here, $\eta_{{c,0}}$ and $\eta_{PC}$ are the coupling efficiencies into the objective lens of the PL emissions coupled with the cavity mode and averaged PPC leaky modes. 
Due to the primarily linear polarization dependence of the cavity modes 1 and 2, the PL spectra shown in Fig. 3(a) with polarizations as   $\phi=0^\circ$ and $90^\circ$ indicate the off- and on-resonance emissions. We calculate the spectrally resolved cavity-enhancement of the collected emission from the two spectra, as shown in Fig. 3(b), which is governed by 
\begin{equation}
\frac{I_{90^\circ}(\lambda)}{I_{0^\circ}(\lambda)}=\frac{\eta_{c,0}}{\eta_{PC}}\frac{
\int\mathrm{d}\mathbf{\upmu}\mathrm{d}^2\mathbf{r}F_{c,0}L(\lambda) |\psi|^{2}\rho(\mathbf{r}, \lambda, \mathbf{\upmu})}{\int\mathrm{d}\mathbf{\upmu}\mathrm{d}^2\mathbf{r}F_{PC}\rho(\mathbf{r}, \lambda, \mathbf{\upmu})}+1
\end{equation}

By integrating the far-field radiations of a dipole spectrally on- or slightly off-resonance with the cavity mode over the numerical aperture (NA=0.95) of the objective lens, which locates on the cavity defect, we obtain the  coupling efficiencies ratios $\frac{\eta_{c,0}}{\eta_{PC}}$ for modes 1 and 2 are 87\% and 73\%, respectively \cite{Gan2013c}. The integral over the angle of the dipole $\mathbf{\upmu}$ with respective to the cavity field  $\mathbf{E}$ equals to $1/2$ due to the random orientations of  dipoles on the two-dimensional MoS$_2$ sheet. The spatial density of the dipoles corresponds to the excitation of a uniform MoS$_2$ area by a Gaussian beam with a full width at half maximum of about 400~nm in the $x-y$ plane. Over this excitation area, the spatial integral of $(|\mathbf{E}||\mathbf{\upmu}|/|\mathbf{E}_{max}||\mathbf{\upmu}|)^2$  are 0.169 and 0.079 for modes 1 and 2, respectively, as calculated from their simulated cavity fields. With the calculated $V_{mode}$ and the $Q$ factors derived from the experimental spectra, we calculate the maximum Purcell factor $F_{c,0}$ for modes 1 and 2 are about 26.5 and 73.8. The suppression factor $F_{PC}$ is estimated by simulating the emission power ratio of a dipole on the L3 cavity defect and on the bulk membrane \cite{2005.PRL.Englund}. The emission frequency of the dipole is chosen in the photonic bandgap of PPC but off-resoance with the cavity mode.  The obtained $F_{PC}$ is approximately 0.4, which is close to the values found in similar PPC structures \cite{2005.PRL.Englund,2005.Science.Noda.SE_control,2005.APL.Fushman.PbS}. Combining the above calculations and the Lorentizan functions $L(\lambda)$ of modes 1 and 2, the theoretical model described in Eq. (3) shows a good fit to the experimentally obtained enhancement spectrum, as shown in Fig. 3(b). 

%   Figure 4(b) shows the spectrally resolved enhancement of the collected emission obtained from the PL spectra with $\phi=90^\circ$ and $0^\circ$ (I_{90^\circ}(\lambda)/I_{0^\circ}(\lambda)). The two peaks are fitted to the Lorentizan cavity modes at  the resonant wavelengthes of 652.6 nm and 656.e nm with the maximum intensities as 5.9 and 6.4, respectively. 
%
%To study the cavity-enhanced SE rate of the monolayer MoS$_2$, we calculate the raito of the PL spectra obtained from the cavity defect with  $\phi=90^\circ$ and $0^\circ$ , as shown in Fig. 4(b). Due to the 
%Since the cavity modes are primarily linearly polarized, the PL spectra obtained with different polarizations indicate the cavity-enhanced PL external extracted efficiency. Figure 4(a) displays the spectrally resolved ratio of the PL with $\phi=90^\circ$ and $0^\circ$ shown in Fig. 2(c). We observe two enhanced peaks, which can be fitted to the Lorentizan resonant modes at the resonant wavelengthes of 652.6 nm and 656.e nm with the maximum intensities as 5.9 and 6.4, respectively. The cavity-enhanced PL external extracted efficiency can be explained by the Purcell enhancement of the SE rate and the enhancement of light colleciton from the cavity mode compared to collected emission from the bulk memebrane.

In conclusion, we have shown that by coupling monolayer MoS$_2$ to an PPC nanocavity, it is possible to dramatically enhance its internal quantum efficiency for transitions on resonant with the cavity. The experimental results and theoretical calculations reveal that the maximum enhancement of the MoS$_2$ SE rate by the cavity modes can be higher than 70, with a suppression factor of about 0.4 due to the PPC lattice.  In this work, the strong Purcell enhancement was limited to the sub-wavelength size of the cavity; however, a high Purcell enhancement across a larger area could be realized using slow light near the bandedge of photonic crystals or coupled cavity arrays~\cite{Altug2005}. The cavity-enhanced light-matter coupling in monolayer MoS$_2$ indicated by the strong Purcell effect expands the scope of solid state cavity quantum electrodynamics to atomically thin materials with large bandgaps, which has implications for a range of optical devices, including efficient photodetectors~\cite{Lopez-Sanchez2013} and electroluminescent systems, cavity-enhanced nonlinearities~\cite{Kumar2013} and potentially even lasers employing atomically thin gain media.

Acknowledgment: The authors thank Kangmook Lim and Edo Waks for the dry etching of PPC cavities. Financial support was provided by the Air Force Office of Scientific Research PECASE, supervised by Dr. Gernot Pom- renke, the DARPA Information in a Photon program, through Grant No. W911NF-10-1-0416. Device fabrication was partly carried out at the Center for Functional Nanoma- terials, Brookhaven National Laboratory, which is supported by the U.S. Department of Energy, Office of Basic Energy Sciences, under Contract No. DE-AC02-98CH10886. R.S. was supported in part by the Center for Excitonics, an Energy Frontier Research Center funded by the U.S. Department of Energy, Office of Science, Office of Basic Energy Sciences under Award Number DE-SC0001088. X.G. was partially supported by the 973 program (2012CB921900) and NSFC (61377035).

%\bibliographystyle{unsrt}
%
%\bibliography{/Users/xuetaogan/Documents/library}

\begin{thebibliography}{99}

\bibitem{Wang2012o} Q. H. Wang, K. Kalantar-Zadeh, A. Kis, J. N. Coleman, and M. S. Strano, Nature Nanotech. {\bf 7}, 699 (2012). 
\bibitem{Mak2010b} K. F. Mak, C. Lee, J. Hone, J. Shan, and T. F. Heinz, Phys. Rev. Lett. {\bf 105}, 136805 (2010).
\bibitem{Han2011a} S. W. Han, H. Kwon, S. K. Kim, S. Ryu, W. S. Yun, D. H. Kim, J. H. Hwang, J. S. Kang, J. Baik, H. J. Shin, and S. C. Hong, Phys. Rev. B {\bf 84}, 045409 (2011).
\bibitem{Kadantsev2012} E. S. Kadantsev and P. Hawrylak, Solid State Commun. {\bf 152}, 909 (2012). 
\bibitem{Radisavljevic2011} B. Radisavljevic, A. Radenovic, J. Brivio, V. Giacometti, and A. Kis, Nature Nanotech. {\bf 6}, 147 (2011).
\bibitem{Splendiani2010} A. Splendiani, L. Sun, Y. Zhang, T. Li, J. Kim, C. Chim, G. Galli, and F. Wang, Nano Lett. {\bf 10}, 1271 (2010).
\bibitem{Korn2012} T. Korn, S. Heydrich, M. Hirmer, J. Schmutzler, C. Schuller, and C. Schu, Appl. Phys. Lett. {\bf 99}, 102109 (2011).
\bibitem{Gan2012b} X. Gan, N. Pervez, I. Kymissis, F. Hatami, and D. Englund, Appl. Phys. Lett. {\bf 100}, 231104 (2012).
\bibitem{Noda2003Nature} Y. Akahane, T. Asano, B. S. Song, and S. Noda, Nature {\bf 425}, 944 (2003).
\bibitem{Mos2010} C. Lee, H. Yan, L. Brus, T. F. Heinz, J. Hone, and S. Ryu, ACS Nano {\bf 4}, 2695 (2010).
\bibitem{2010.NNano.Hone.graphene} C. R. Dean, A. F. Young, I. Meric, C. Lee, L. Wang, S. Sorgenfrei, K. Watanabe, T. Taniguchi, P. Kim, K. L. Shepard, and J. Hone, Nature Nanotech. {\bf 5}, 722 (2010).
\bibitem{2005.Science.Noda.SE_control} M. Fujita, S. Takahashi, Y. Tanaka, T. Asano, and S. Noda, Science {\bf 308}, 1296 (2005).
\bibitem{2007.APL.UK_ppl.mode_structure_L3} A. R. A. Chalcraft, S. Lam, D. OBrien, T. F. Krauss, M. Sahin, D. Szymanski, D. Sanvitto, R. Oulton, M. S. Skolnick, A. M. Fox, D. M. Whittaker, H. Y. Liu, and M. Hopkinson, Appl. Phys. Lett. {\bf 90}, 241117 (2007).
\bibitem{Schwagmann2012} A. Schwagmann, S. Kalliakos, D. J. P. Ellis, I. Farrer, J. P. Griffiths, G. C. Jones, D. Ritchie, and A. J. Shields, Opt. Express {\bf 20}, 28614 (2012).
\bibitem{2005.PRL.Englund} D. Englund, D. Fattal, E. Waks, G. Solomon, B. Zhang, T. Nakaoka, Y. Arakawa, Y. Yamamoto, and J. Vuckovic, Phys. Rev. Lett. {\bf 95}, 13904 (2005).
\bibitem{Kress2005} A. Kress, F. Hofbauer, N. Reinelt, M. Kaniber, H. Krenner, R. Meyer, G. B?ohm, and J. Finley, Phys. Rev. B {\bf 71}, 241304 (2005).
\bibitem{Kaniber2007} M. Kaniber, A. Kress, A. Laucht, M. Bichler, R. Meyer, M. C. Amann, and J. J. Finley, Appl. Phys. Lett. {\bf 91}, 061106 (2007).
\bibitem{Gan2012} X. Gan, K. F. Mak, Y. Gao, Y. You, F. Hatami, J. Hone, T. F. Heinz, and D. Englund, Nano Lett. {\bf 12}, 5626 (2012).
\bibitem{Gan2013b} X. Gan, R. Shiue, Y. Gao, K. F. Mak, X. Yao, L. Li, A. Szep, D. Walker, J. Hone, T. F. Heinz, and D. Englund, Nano Lett. {\bf 13}, 691 (2013).
\bibitem{Gan2013c} X. Gan, H. Clevenson, C. Tsai, L. Li, and D. Englund, Sci. Rep. {\bf 3}, 2145 (2013).
\bibitem{2005.APL.Fushman.PbS} I. Fushman, D. Englund, and J. Vuckovic, Appl. Phys. Lett. {\bf 87}, 241102 (2005).
\bibitem{Altug2005} H. Altug and J. Vuckovic, Appl. Phys. Lett. {\bf 86}, 111102 (2005).
\bibitem{Lopez-Sanchez2013} O. Lopez-Sanchez, D. Lembke, M. Kayci, A. Radenovic, and A. Kis, Nature Nanotech. {\bf 8}, 497 (2013).
\bibitem{Kumar2013} N. Kumar, S. Najmaei, Q. Cui, F. Ceballos, P. M. Ajayan, J. Lou, and H. Zhao, Phys. Rev. B {\bf 87}, 161403 (2013).

\end{thebibliography}

\end{document}